\documentclass[prd,amsmath,amssymb,showpacs,superscriptaddress,nofootinbib,twocolumn]{revtex4-1}
\usepackage{multirow}
\usepackage{epsfig}
\usepackage{dcolumn}
\usepackage{bm}
\usepackage{color}
\usepackage{graphicx}
\usepackage{subfigure}
\usepackage{pstricks}
\usepackage{pst-node}
\usepackage{rotating}
\usepackage{times}
\usepackage{overpic}
\usepackage[colorlinks,linkcolor=blue,citecolor=red]{hyperref}
\usepackage{lineno}
\usepackage{subfigure}
\usepackage{footnote}
\usepackage[english]{babel}
\addto{\captionsenglish}{%

}

\newcommand{\BR}{\mathcal{B}}
\newcommand{\fz}{f_{0}(980)}
\newcommand{\fone}{f_{1}(1285)}
\newcommand{\jpsi}{J/\psi}
\newcommand{\piz}{\pi^{0}}
\newcommand{\pip}{\pi^{+}}
\newcommand{\pim}{\pi^{-}}
\newcommand{\Kp}{K^{+}}
\newcommand{\Km}{K^{-}}

\begin{document}
\widetext

\title{\boldmath Observation of the isospin-violating decay $\jpsi \to \phi\pi^{0}f_{0}(980)$
}

\author{
  \begin{small}
    \begin{center}
      M.~Ablikim$^{1}$, M.~N.~Achasov$^{9,a}$, X.~C.~Ai$^{1}$,
      O.~Albayrak$^{5}$, M.~Albrecht$^{4}$, D.~J.~Ambrose$^{44}$,
      A.~Amoroso$^{48A,48C}$, F.~F.~An$^{1}$, Q.~An$^{45}$,
      J.~Z.~Bai$^{1}$
% ...
      \\
      \vspace{0.2cm}
      (BESIII Collaboration)\\
      \vspace{0.2cm} {\it
        $^{1}$ Institute of High Energy Physics, Beijing 100049, People's Republic of China\\
% ...
$^{4}$ Bochum Ruhr-University, D-44780 Bochum, Germany\\
        $^{5}$ Carnegie Mellon University, Pittsburgh, Pennsylvania 15213, USA\\
% ...
$^{9}$ G.I. Budker Institute of Nuclear Physics SB RAS (BINP), Novosibirsk 630090, Russia\\
% ...
$^{44}$ University of Rochester, Rochester, New York 14627, USA\\
        $^{45}$ University of Science and Technology of China, Hefei 230026, People's Republic of China\\
% ...
$^{48}$ (A)University of Turin, I-10125, Turin, Italy; (B)University of Eastern Piedmont, I-15121, Alessandria, Italy; (C)INFN, I-10125, Turin, Italy\\
% ...
$^{a}$ Also at the Novosibirsk State University, Novosibirsk, 630090, Russia\\
% ...
      }\end{center} \vspace{0.4cm}
  \end{small}
}
\affiliation{}


\begin{abstract}
  Using a sample of 1.31 billion $\jpsi$ events collected with the
  BESIII detector at the BEPCII collider, the decays
  $\jpsi \to \phi \pip\pim\piz$ and $\jpsi \to \phi \piz\piz\piz$ are
  investigated. The isospin violating decay $\jpsi \to \phi \piz \fz$
  with $\fz \to \pi\pi$, is observed for the first time. The width of
  the $\fz$ obtained from the dipion mass spectrum is found to be much
  smaller than the world average value. In the $\piz \fz$ mass
  spectrum, there is evidence of $\fone$ production. By studying the
  decay $\jpsi \to \phi\eta'$, the branching fractions of
  $\eta' \to \pip\pim\piz$ and $\eta' \to \piz\piz\piz$, as well as their ratio, are also
  measured.
\end{abstract}

\maketitle

\section{Introduction}
The nature of the scalar meson $\fz$ is a long-standing puzzle. It has
been interpreted as a $q\bar{q}$ state, a $K\bar{K}$ molecule, a
glueball, and a four-quark state (see the review in
Ref.~\cite{PDG14}). Further insights are expected from studies of
$\fz$ mixing with the $a_0^0(980)$~\cite{a0f0mix}, evidence for which
was found in a recent BESIII analysis of $J/\psi$ and $\chi_{c1}$
decays~\cite{BESa0f0}. BESIII also observed a large isospin violation
in $\jpsi$ radiatively decaying into $\pip\pim\piz$ and $\piz\piz\piz$
involving the intermediate decay $\eta(1405) \to \piz
\fz$~\cite{BESgpi0f0}. In this study, the $\fz$ width was found to be
$9.5\pm1.1$~MeV/$c^2$. One proposed explanation for this anomalously
narrow width and the observed large isospin violation, which cannot be caused
by $a_0^0(980)-f_0(980)$ mixing, is the triangle singularity
mechanism~\cite{JJWu, Aceti}.

The decays $\jpsi \to \phi \pip\pim\piz$ and $\jpsi \to \phi
\piz\piz\piz$ are similar to the radiative decays $\jpsi \to \gamma
\pip\pim\piz/\piz\piz\piz$ as the $\phi$ and $\gamma$ share the same
spin and parity quantum numbers. Any intermediate $\fz$ would be
noticeable in the $\pi\pi$ mass spectra. At the same time, a study of the
decay $\jpsi \to \phi \eta'$ would enable a measurement of the
branching fractions for $\eta' \to \pip\pim\piz$ and $\eta' \to
\piz\piz\piz$. The recently measured $\BR(\eta' \to 3\piz) =
(3.56\pm0.40)\times 10^{-3}$~\cite{BESgpi0f0} from a study of the
decay $\jpsi \to \gamma \eta'$ was found to be nearly $4\sigma$ higher
than the previous value $(1.73\pm0.23)\times 10^{-3}$ from studies of
the reaction $\pim p \to n (6\gamma)$~\cite{etap3piz1,etap3piz2,etap3piz3}\footnote{The PDG~\cite{PDG14}
  gives an average value, $\Gamma(\eta' \to 3\piz)/\Gamma(\eta' \to
  \piz\piz\eta) = 0.0078 \pm 0.0010$, of three
  measurements~\cite{etap3piz1,etap3piz2,etap3piz3}. $\BR(\eta'
  \to 3\piz)$ is calculated using $\BR(\eta' \to \piz\piz\eta) = 0.222
  \pm 0.008$~\cite{PDG14}, assuming the uncertainties are
  independent.}.
Additionally, the isospin-violating decays $\eta' \to
\pip\pim\piz/\piz\piz\piz$ provide a means to extract the $d$, $u$
quark mass difference $m_d-m_u$~\cite{Gross}.

This paper reports a study of $\jpsi \to \phi \pip\pim\piz$ and $\jpsi
\to \phi \piz\piz\piz$ with $\phi \to \Kp\Km$ based on a sample of
$(1.311 \pm 0.011) \times 10^{9}$~\cite{Njpsi09, Njpsi12} $\jpsi$
events accumulated with the BESIII detector in 2009 and 2012.

\section{Detector and Monte Carlo Simulation}
The BESIII detector~\cite{BES} is a magnetic spectrometer located at
the Beijing Electron-Positron Collider (BEPCII), which is a
double-ring $e^{+}e^{-}$ collider with a design luminosity of
$10^{33}$ cm$^{-2}$s$^{-1}$ at a center of mass (c.m.) energy of
3.773~GeV. The cylindrical core of the BESIII detector consists of a
helium-based main drift chamber (MDC), a plastic scintillator
time-of-flight system (TOF), and a CsI(Tl) electromagnetic calorimeter
(EMC). All are enclosed in a superconducting solenoidal magnet
providing a 1.0 T (0.9 T in 2012) magnetic field. The solenoid is
supported by an octagonal flux-return yoke with resistive plate
counter muon identifier modules interleaved with steel. The acceptance
for charged tracks and photons is $93\%$ of $4\pi$ solid angle. The
charged-particle momentum resolution is $0.5\%$ at 1~GeV/$c$, and the
specific energy loss ($dE/dx$) resolution is better than $6\%$. The
photon energy is measured in the EMC with a resolution of $2.5\%$
($5\%$) at 1~GeV in the barrel (endcaps). The time resolution of the
TOF is 80 ps (110 ps) in the barrel (endcaps). The BESIII offline
software system framework, based on the \textsc{Gaudi} package~\cite{Gaudi},
provides standard interfaces and utilities for event simulation, data
processing and physics analysis.

Monte Carlo (MC) simulation, based on the \textsc{GEANT4}~\cite{Geant4}
package, is used to simulate the detector response, study the
background and determine efficiencies. For this analysis, we use a
phase space MC sample to describe the three body decay $\jpsi \to \phi
\piz \fz$, while the angular distributions are considered in the
decays $\jpsi \to \phi \fone \to \phi \piz\fz$ and $\jpsi \to
\phi\eta'$.
In the MC samples, the width of the $\fz$ is fixed to be
$15.3$~MeV/$c^2$, which is obtained from a fit to data as described
below.
An inclusive MC sample of 1.2 billion $\jpsi$ decays is used to study
the background. For this MC sample, the generator
\textsc{BesEvtGen}~\cite{EvtGen, BesMC} is used to generate the known $\jpsi$
decays according to their measured branching fractions~\cite{PDG14}
while \textsc{Lundcharm}~\cite{Lund} is used to generate the remaining unknown
decays.

\section{Event Selection}
Charged tracks are reconstructed from hits in the MDC and selected by requiring that $|\cos\theta|<0.93$,
where $\theta$ is the polar angle measured in the MDC, and that the
point of closest approach to the $e^+e^-$ interaction point is within
$\pm 10$~cm in the beam direction and within 1~cm in the plane
perpendicular to the beam direction. TOF and $dE/dx$ information are
combined to calculate the particle identification (PID) probabilities
for the pion, kaon and proton hypotheses. For each photon, the energy
deposited in the EMC must be at least 25~MeV (50~MeV) in the region of
$|\cos\theta|<0.8$ ($0.86<|\cos\theta|<0.92$). To exclude showers that
originate from charged tracks, the angle between a photon candidate
and the closest charged track must be larger than $10^\circ$. The
timing information from the EMC is used to suppress electronics noise
and unrelated energy deposits.

To be accepted as a $\jpsi \to \Kp\Km\pip\pim \piz$ decay, a candidate
event is required to have four charged tracks with zero net charge and
at least two photons. The two oppositely charged tracks with an
invariant mass closest to the nominal mass of the $\phi$ are assigned
as being kaons, while the remaining tracks are assigned as being
pions. To avoid misidentification, kaon tracks are required to have a
PID probability of being a kaon that is larger than that of being a
pion. A 5-constraint kinematic fit is applied to the candidate events
under the hypothesis $\jpsi \to \Kp\Km\pip\pim\gamma\gamma$. This
includes a constraint that the total four-momenta of the selected
particles must be equal to the initial four-momentum of the colliding
beams (4-constraint) and that the invariant mass of the two photons
must be the nominal mass of the $\piz$ (1-constraint). If
more than 2 photon candidates are found in the event,
the combination with the minimum
$\chi^{2}(5C)$ from the kinematic fit is retained. Only events with a
$\chi^{2}(5C)$ less than 100 are accepted. Events with a
$K^{\pm}\pi^{\mp}$ invariant mass satisfying
$|M(K^{\pm}\pi^{\mp})-M(K^{*0})|<0.050$~GeV/$c^2$ are rejected in order
to suppress the background containing $K^{*0}$ or $\bar{K}^{*0}$
intermediate states.

To be accepted as a $\jpsi \to \Kp\Km\piz\piz\piz$ decay, a candidate
event is required to have two oppositely charged tracks and at least
six photons. For both tracks, the PID probability of being a kaon must
be larger than that of being a pion.
The six photons are selected and paired by minimizing the quantity
$\frac{(M(\gamma_1\gamma_2)-M_{\piz})^2}{\sigma_{\piz}^2} +
\frac{(M(\gamma_3\gamma_4)-M_{\piz})^2}{\sigma_{\piz}^2} +
\frac{(M(\gamma_5\gamma_6)-M_{\piz})^2}{\sigma_{\piz}^2}$,
where $M(\gamma_i\gamma_j)$ is the mass of $\gamma_i$$\gamma_j$, and
$M_{\piz}$ and $\sigma_{\piz}$ are the nominal mass and reconstruction
resolution of the $\piz$ respectively. A 7-constraint kinematic fit is
performed to the $\jpsi \to \Kp\Km 6\gamma$ hypothesis, where the
constraints include the four-momentum constraint to the four-momentum of the
colliding beams and three constraints of photon pairs to have
invariant masses equal to the $\piz$. Events with a $\chi^{2}(7C)$
less than 90 are accepted.

\begin{figure}
  \subfigure{\includegraphics[width=0.4\textwidth]{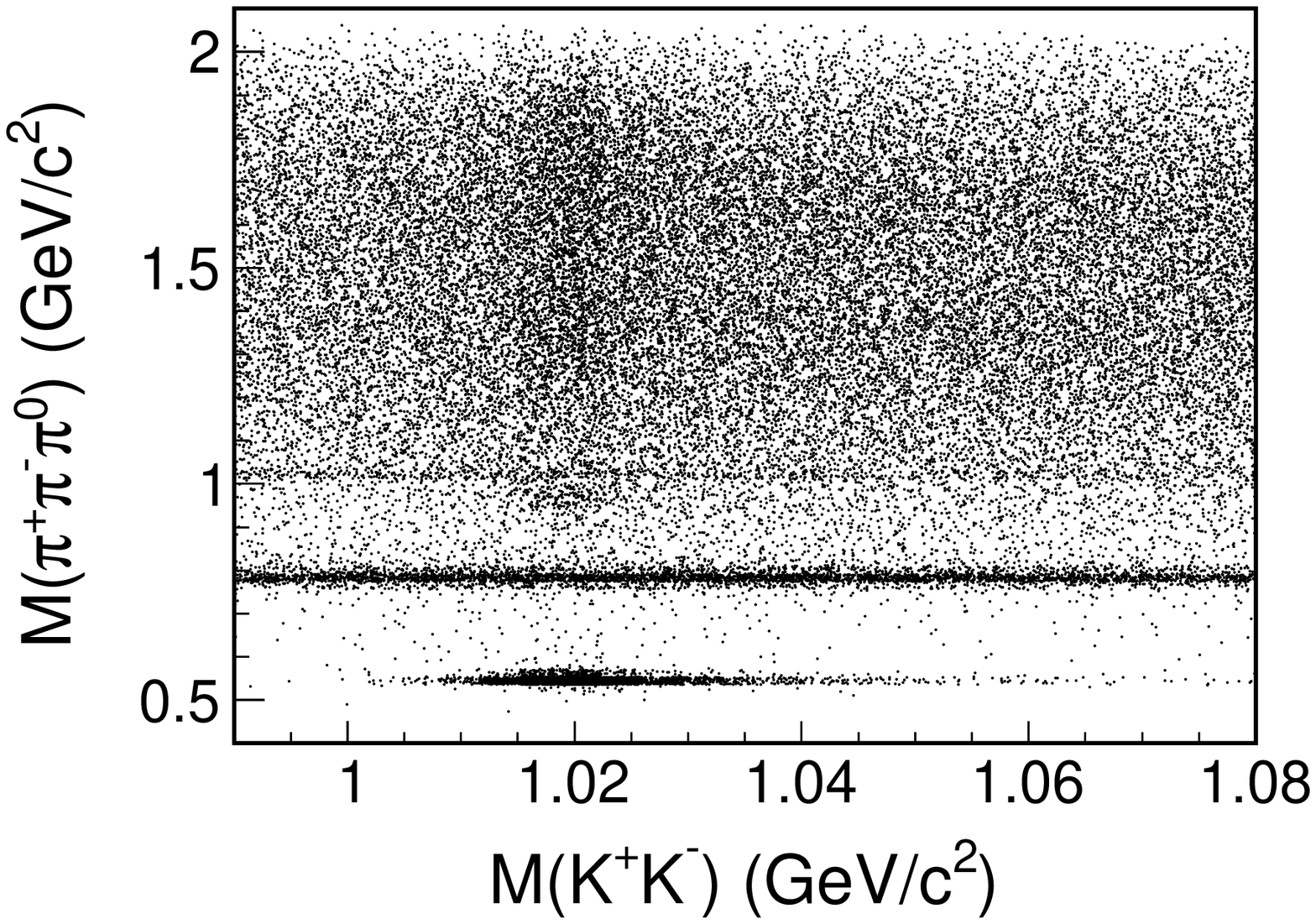}}
  \put(-155, 120){\textbf{(a)}}\\
  \subfigure{\includegraphics[width=0.4\textwidth]{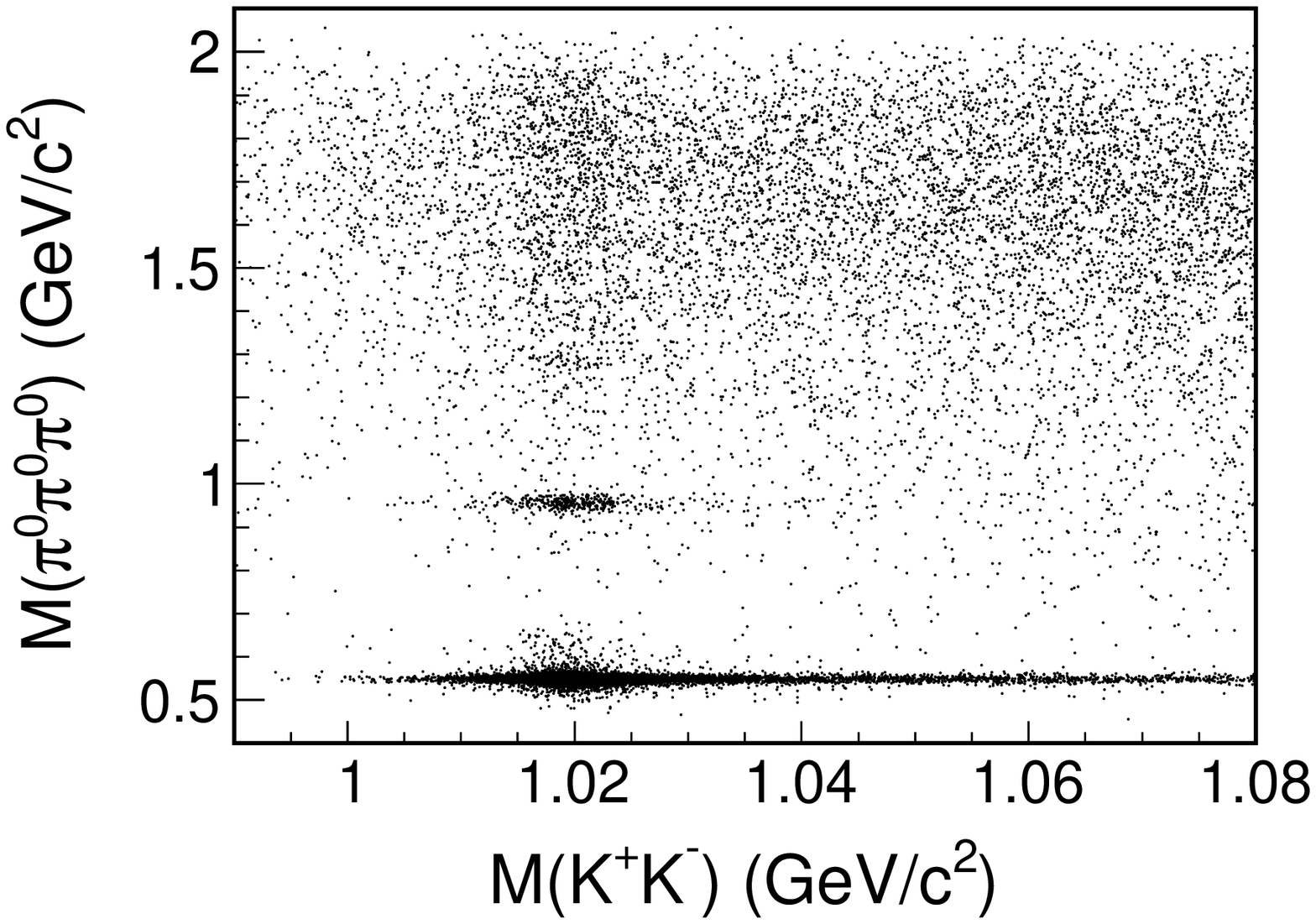}}
  \put(-155, 120){\textbf{(b)}}\\
  \caption{Scatter plots of (a) $M(\pip\pim\piz)$ versus $M(\Kp\Km)$ and (b) $M(\piz\piz\piz)$ versus $M(\Kp\Km)$.}\label{fig:Mf0pizMf0}
\end{figure}

Figures~\ref{fig:Mf0pizMf0} (a) and (b) show $M(3\pi)$ versus
$M(\Kp\Km)$ for the two final states respectively. Clear signals from
$\phi\eta$ and $\phi\eta'$ with $\eta' \to 3\piz$ are noticeable. In
Fig.~\ref{fig:Mf0pizMf0} (a), horizontal bands are noticeable from
$\omega$ and $\phi$ decaying into $\pip\pim\piz$ in the background channel
$\jpsi \to \omega/\phi \Kp\Km$.

To search for the decay $\jpsi \to \phi \piz \fz$, we focus on the region $0.99<M(\Kp\Km)<1.06$~GeV/$c^2$ and $0.850<M(\pi\pi)<1.150$~GeV/$c^2$. The $M(\Kp\Km)$ spectra are shown in Fig.~\ref{fig:fitphi}. Clear $\phi$ signals are visible.
The $M(\pip\pim)$ and $M(\piz\piz)$ spectra for the $\phi$ signal
region, which is defined by requiring
$1.015<M(\Kp\Km)<1.025$~GeV/$c^2$, are presented in
Fig.~\ref{fig:fitf0} (a) and (b) respectively. A clear $\fz$ peak
exists for the $\pip\pim$ mode.
The $M(\fz[\pi\pi]\piz)$ spectra for the $\fz$ signal region, defined
as $0.960<M(\pi\pi)<1.020$~GeV/$c^2$, are presented in
Fig.~\ref{fig:fitf1}. There is evidence of a resonance around
1.28~GeV/$c^2$ for the decay $\fz \to \pip\pim$, which will be identified as the $\fone$~\footnote{For
  simplicity, $\fz$ and $\fone$ will be written as $f_0$ and $f_1$
  respectively throughout this paper.}.

To ensure that the observed $f_0$ and $f_1$ signals do not originate
from background processes, the same selection criteria as described
above are applied to an MC sample of $1.2$ billion inclusive $\jpsi$
decays which does not contain the signal decay. As expected,
neither an $f_1$ nor an $f_0$ is observed from the inclusive MC
sample. The non-$\phi$ background is studied using data from the
$\phi$ sideband regions ($0.990<M(\Kp\Km)<1.000$~GeV/$c^2$ and
$1.040<M(\Kp\Km)<1.050$~GeV/$c^2$), which are given by the hatched
histograms in Fig.~\ref{fig:fitf0} and Fig.~\ref{fig:fitf1} and in
which no $f_0$ or $f_1$ signals are observed.

\section{Signal extraction of $\jpsi \to \phi\piz\fz$}
\begin{figure}
  \subfigure{\includegraphics[width=0.4\textwidth]{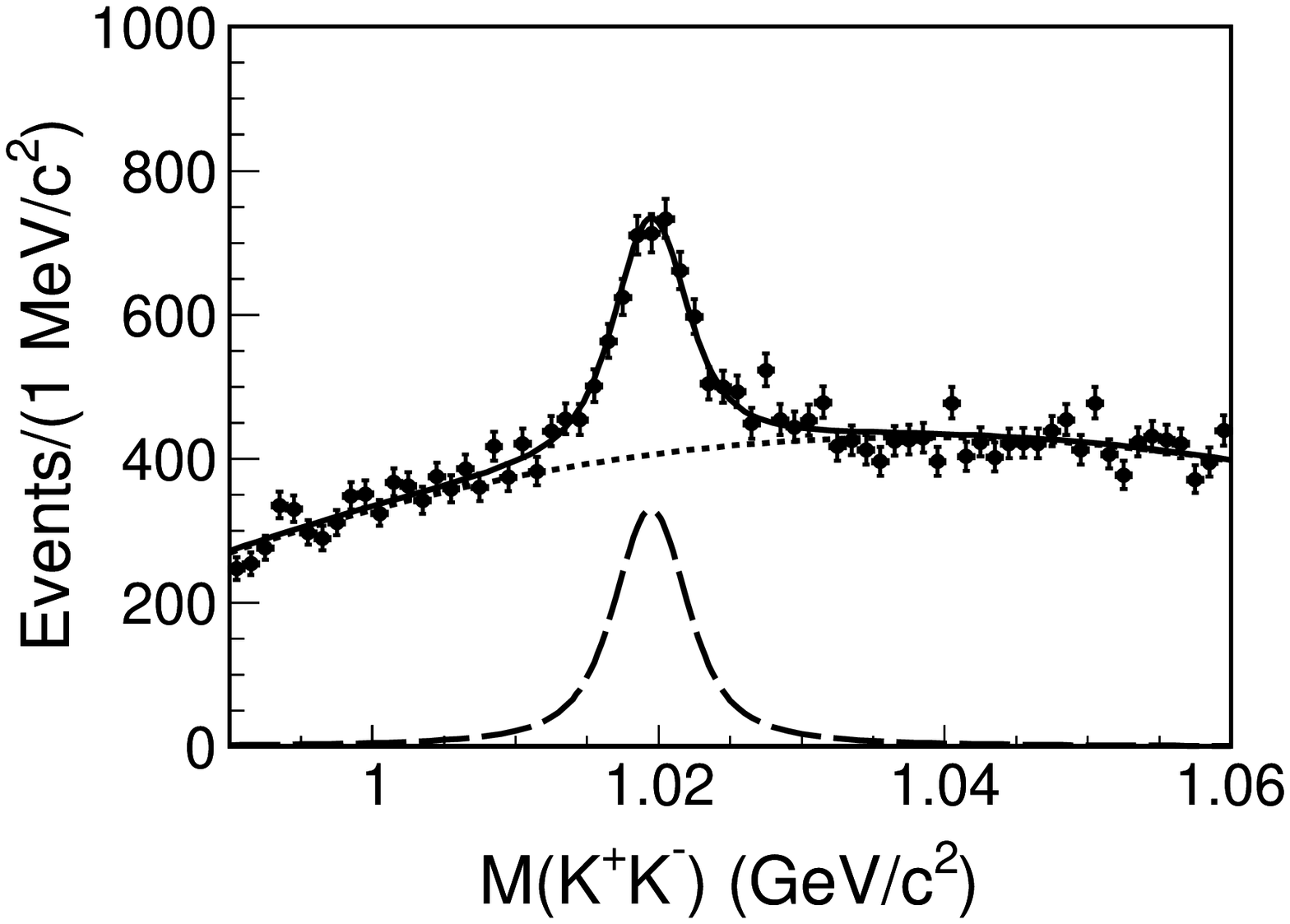}}
  \put(-155, 120){\textbf{(a)}}\\
  \subfigure{\includegraphics[width=0.4\textwidth]{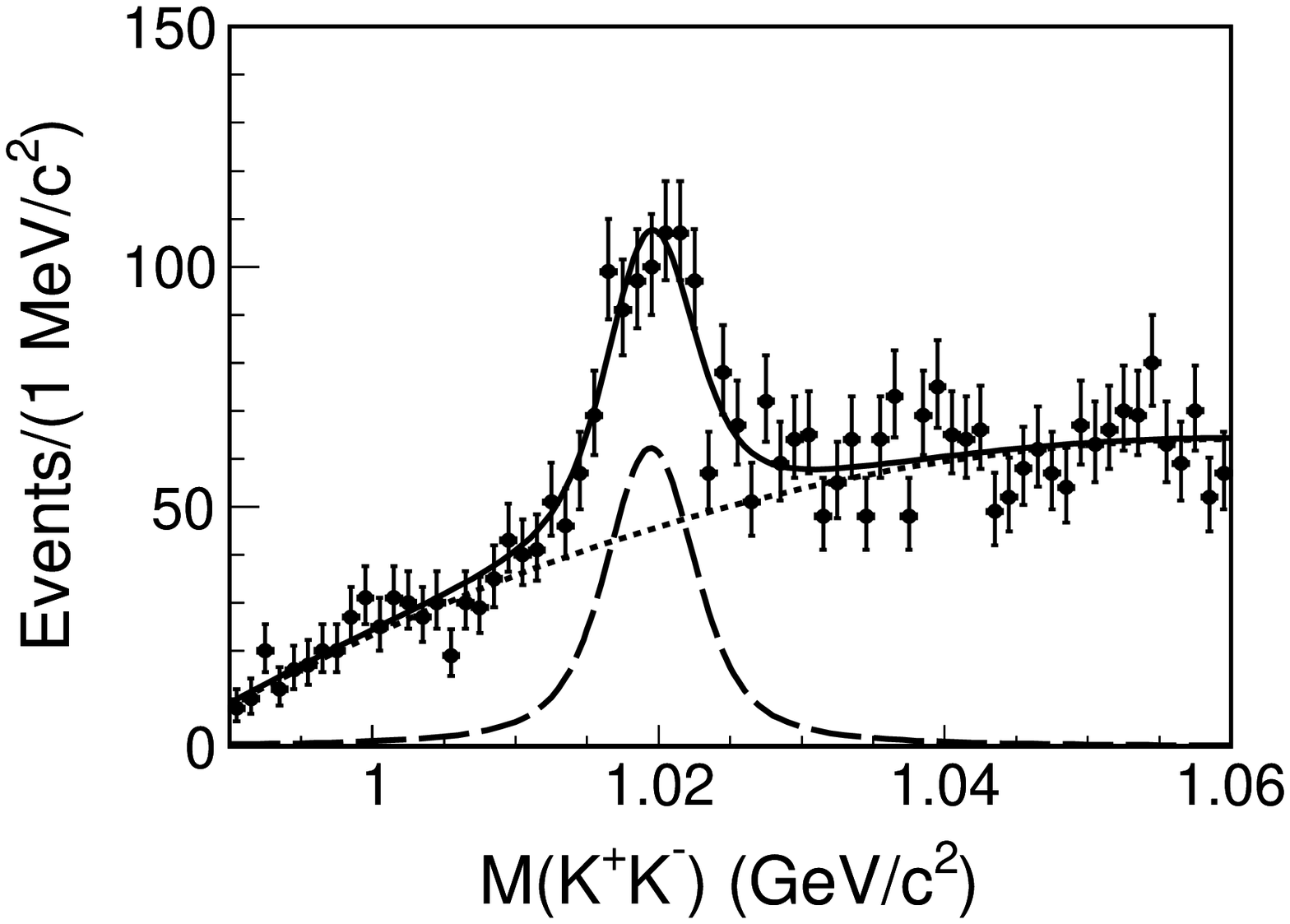}}
  \put(-155, 120){\textbf{(b)}}\\
  \caption{\label{fig:fitphi} Fits to the $M(\Kp\Km)$ mass spectra for
    the mode (a) $\fz \to \pip\pim$ and (b) $\fz \to \piz\piz$. The
    solid curve is the full fit; the long-dashed curve is the $\phi$
    signal while the short-dashed curve is the background.}
\end{figure}

\begin{figure}
  \subfigure{\includegraphics[width=0.4\textwidth]{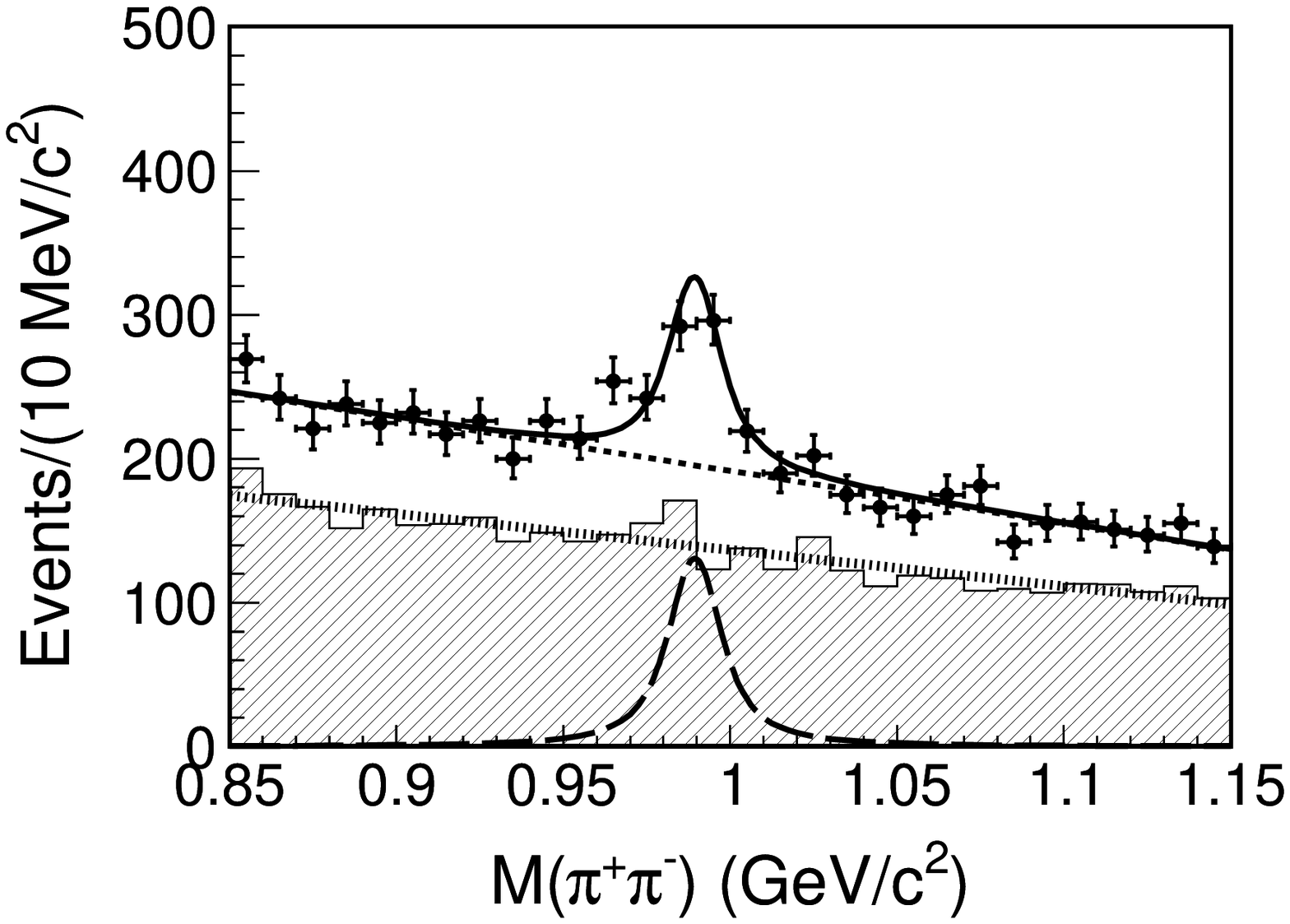}}
  \put(-155, 120){\textbf{(a)}}\\
  \subfigure{\includegraphics[width=0.4\textwidth]{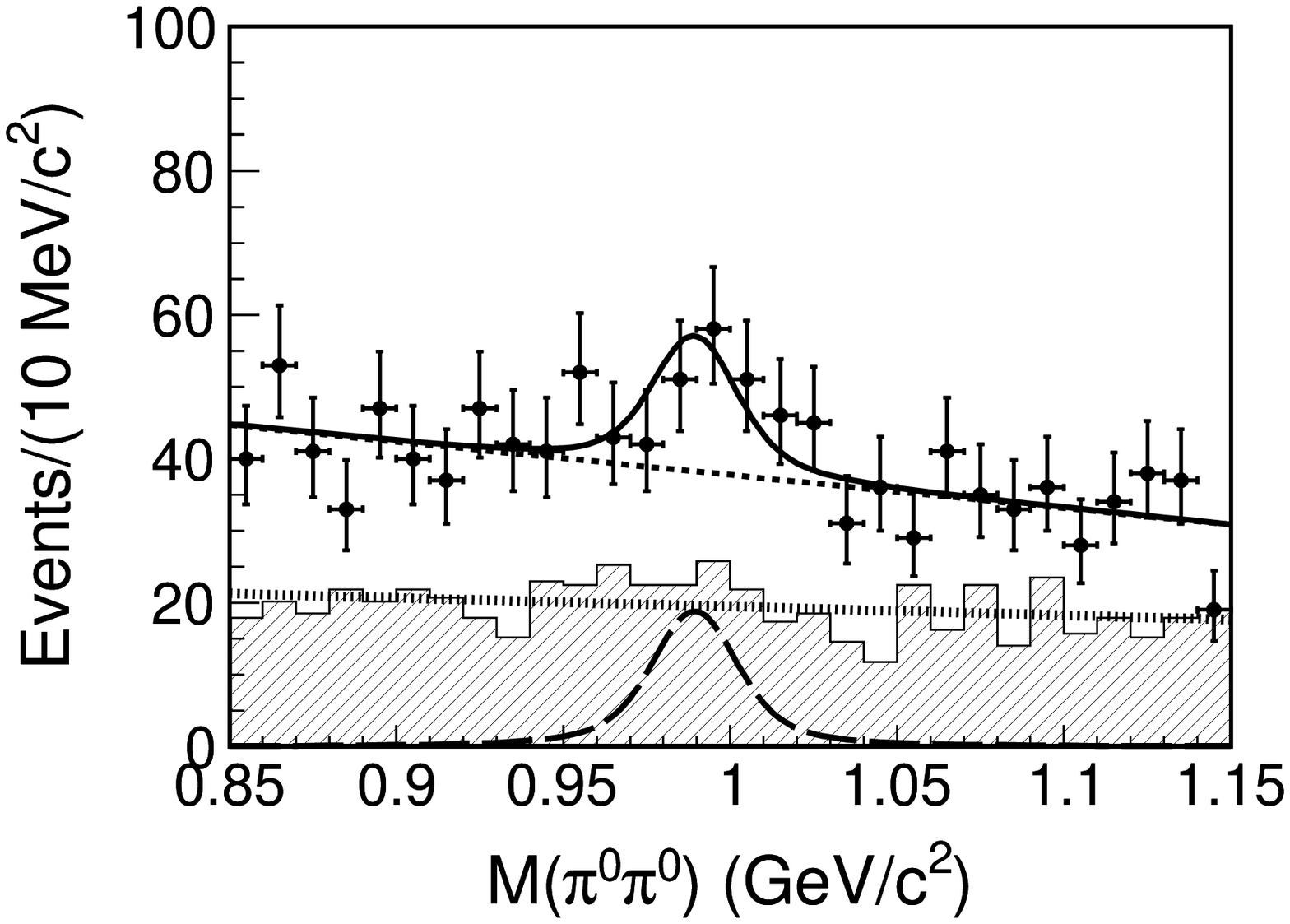}}
  \put(-155, 120){\textbf{(b)}}\\
  \caption{\label{fig:fitf0}The spectra (a) $M(\pip\pim)$ and (b)
    $M(\piz\piz)$ (three entries per event) with $\Kp\Km$ in the
    $\phi$ signal region (the black dots) and in the $\phi$ sideband
    regions (the hatched histogram). The solid curve is the full fit;
    the long-dashed curve is the $\fz$ signal; the dotted line is the
    non-$\phi$ background and the short-dashed line is the total
    background. }
\end{figure}

Figures~\ref{fig:fitf0} (a) and (b) show the $\pip\pim$ and $\piz\piz$
mass spectra for events with $M(\Kp\Km)$ in the $\phi$ signal region
(the black dots) and sideband regions (the hatched histogram scaled by
a normalization factor, $C$). Events in the $\phi$ sideband regions
are normalized in the following way. A fit is performed to the
$\Kp\Km$ mass spectrum, where the $\phi$ signal is described by a
Breit-Wigner function convoluted with a Gaussian resolution function
and the background is described by a second-order polynomial. The mass
and width of the $\phi$ resonance are fixed to their world average
values~\cite{PDG14} and the mass resolution is allowed to float. The
normalization factor $C$ is defined as $ A_\text{sig}/A_\text{sbd}$, where
$A_\text{sig}$ ($A_\text{sbd}$) is the area of the background function from the
fits in the signal (sideband) region. The results of the fits are
shown in Fig.~\ref{fig:fitphi} (a) and (b).

To extract the signal yield of $\jpsi \to \phi\piz f_0$, a
simultaneous unbinned maximum likelihood fit is performed to the
$\pip\pim$ and $\piz\piz$ mass spectra. The lineshape of the $f_0$ signal
is different from that of the Flatt\'{e}-form resonance observed in
the decays $\jpsi \to \phi\pip\pim$ and $\jpsi \to
\phi\Kp\Km$~\cite{BESf0}. A Breit-Wigner function convoluted with a
Gaussian mass resolution function is used to describe the $f_0$
signal. The mass resolutions of the $f_0$ in the $M(\pip\pim)$ and
$M(\piz\piz)$ spectra are determined from MC simulations. The
non-$\phi$ background is parameterized with a straight line,
which is determined from a fit to the data in the $\phi$ sideband
regions. The size of this polynomial is fixed according to the
normalized number of background events under the $\phi$ peak,
$N_\text{bkg}=C N_\text{sbd}$, where $N_\text{sbd}$ is the number of events falling
in the $\phi$ sideband regions and $C$ is the normalization factor
obtained above. Another straight line is used to account for
the remaining background from $\jpsi \to \phi \piz \pi\pi$ without
$f_0$ decaying into $\pi\pi$.

The mass and width of the $f_0$ are constrained to be the same for
both the $\Kp\Km\pip\pim\piz$ and the $\Kp\Km\piz\piz\piz$ final
states. The fit yields the values $M(f_0)=989.4\pm1.3$~MeV/$c^2$ and
$\Gamma(f_0)= 15.3 \pm 4.7$~MeV/$c^2$, with the number of events
$N=354.7\pm63.3$ for the $\pip\pim$ mode and $69.8\pm21.1$ for the
$\piz\piz$ mode.
The statistical significance is determined by the changes of the log
likelihood value and the number of degrees of freedom in the fit with
and without the signal~\cite{James}. The significance of the $f_0$
signal is $9.4\sigma$ in the $\Kp\Km\pip\pim\piz$ final state and
$3.2\sigma$ in the $\Kp\Km\piz\piz\piz$ final state. The measured mass
and width obtained from the invariant dipion mass spectrum are
consistent with those from the study of the decay $\jpsi \to
\gamma\eta(1405) \to \gamma \piz\fz$~\cite{BESgpi0f0}. It is worth
noting that the measured width of the $f_0$ observed in the dipion
mass spectrum is much smaller than the world average value of
40-100~MeV~\cite{PDG14}.

\section{Signal extraction of $\jpsi \to \phi \fone$ with $\fone \to \piz \fz$}
\begin{figure}
  \subfigure{\includegraphics[width=0.4\textwidth]{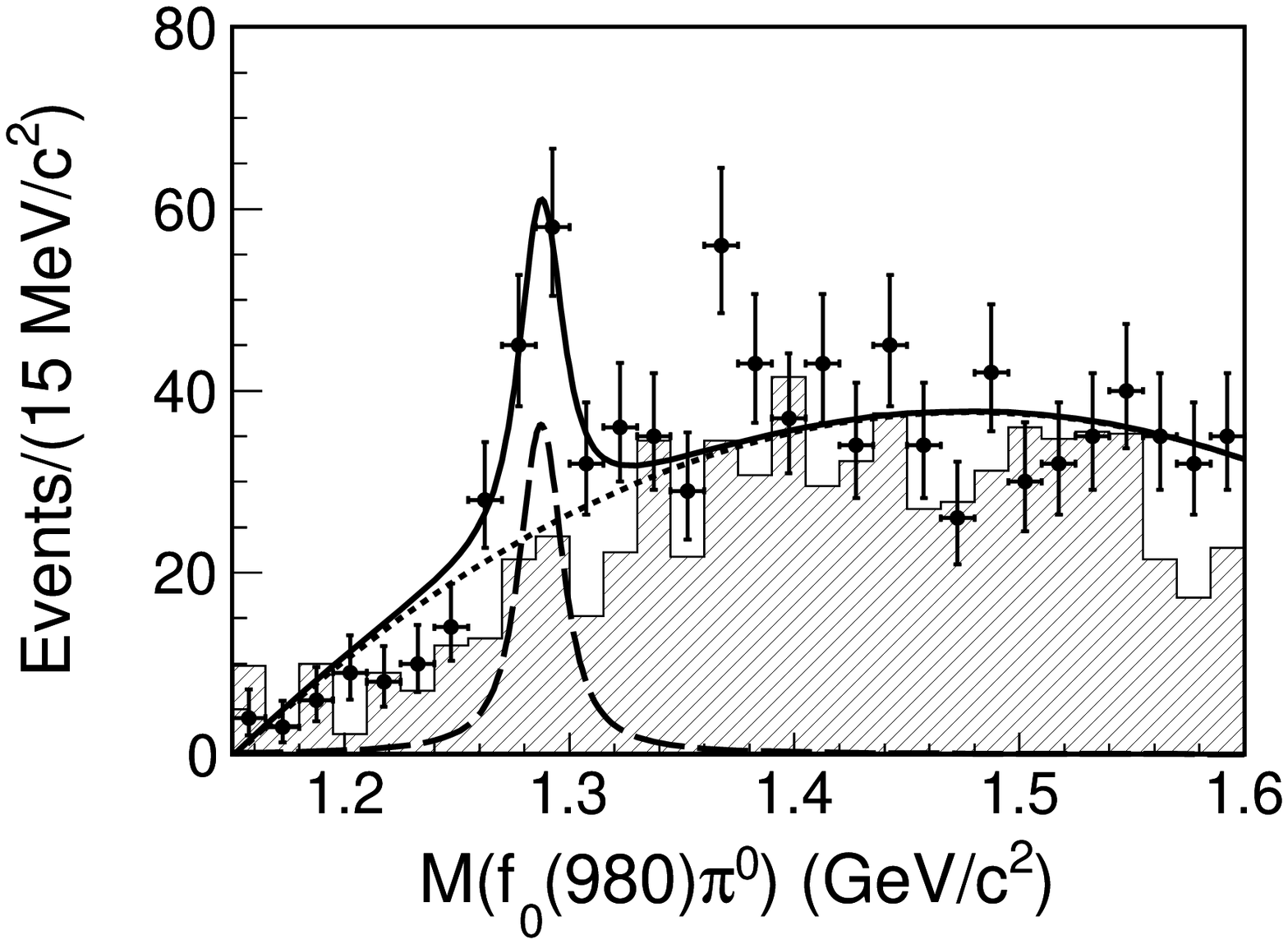}}
  \put(-155, 120){\textbf{(a)}}\\
  \subfigure{\includegraphics[width=0.4\textwidth]{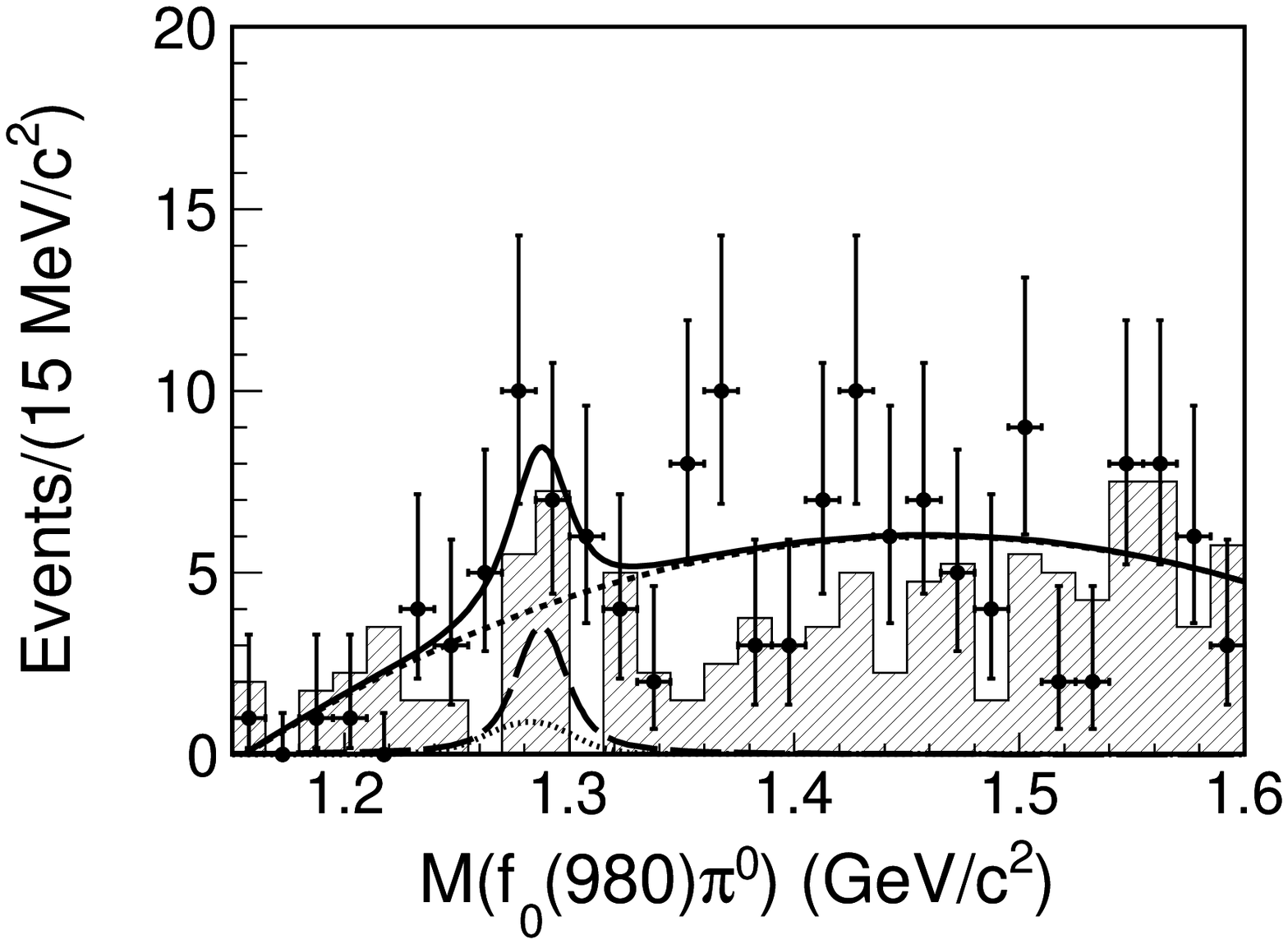}}
  \put(-155, 120){\textbf{(b)}}
  \caption{The spectra of (a) $M(\pip\pim\piz)$ and (b) $M(\piz\piz\piz)$
    in the $\phi$ and $\fz$ signal region (the black dots with error
    bars) and in the sideband regions (the hatched histogram). The
    solid curve is the result of the fit, the long-dashed curve is the
    $\fone$ signal, and the short-dashed curve is the background. In
    (b), the dotted curve represents the peaking background from the
    decay $\fone\to\piz\piz\eta/\piz a_0^0$ with $\eta \to
    \gamma\gamma$.}\label{fig:fitf1}
\end{figure}

Figures~\ref{fig:fitf1} (a) and (b) show the $\pip\pim\piz$ and
$\piz\piz\piz$ mass spectra in the $\phi$ and $f_0$ signal region (the
black dots) and sideband regions (the hatched histogram). The $f_0$
sideband regions are defined as $0.850<M(\pi\pi)<0.910$~GeV/$c^2$ and
$1.070<M(\pi\pi)<1.130$~GeV/$c^2$. In Fig.~\ref{fig:fitf1}, events in
the 2-dimensional sideband regions are weighted as follows. Events
that fall in only the $\phi$ or $\fz$ sideband regions are given a
weight 0.5 to take into account the non-$\phi$ or non-$\fz$ background
while those that fall in both the $\phi$ and the $\fz$ sideband
regions are given a weight $-0.25$ to compensate for the double
counting of the non-$\phi$ and non-$\fz$ background. There is evidence
of a resonance around 1.28~GeV/$c^2$ that is not noticeable in the
2-dimensional sideband regions. By studying an MC sample of the decay
$\jpsi \to \phi f_1 \to \text{anything}$, we find that the decay $f_1 \to
\piz\piz\eta/\piz a_0^0$~\footnote{For simplicity, $a_0(980)$ and
  $a_0^0(980)$ are written as $a_0$ and $a_0^0$ respectively
  throughout this paper.} with $\eta \to \gamma\gamma$ contributes as
a peaking background for the decay $f_1 \to \piz\piz\piz$. The yield
of this peaking background is calculated to be $3.1 \pm 0.6$ using the
relevant branching fractions~\footnote{We assume that $\BR(f_1 \to
  \piz\piz\eta)=\frac{1}{3}\BR(f_1 \to \pi\pi\eta)$, $\BR(f_1 \to \piz
  a_0^0)=\frac{1}{3}\BR(f_1 \to \pi a_0)$, and
  $\BR(a_0^0\to\piz\eta)=100\%$.}~\cite{PDG14} and the efficiency
determined from an MC simulation. A simultaneous unbinned maximum
likelihood fit is performed to the $M(\pip\pim\piz)$ and
$M(\piz\piz\piz)$ distributions. The $f_1$ signal is described by a
Breit-Wigner function convoluted with a Gaussian mass resolution
function. The shape of the peaking background $f_1 \to
\piz\piz\eta/\piz a_0^0$ is determined from an exclusive MC sample and
its size is fixed to be 3.1. A second order polynomial function is
used to describe the remaining background. The mass resolutions of the
$f_1$ in $M(\pip\pim\piz)$ and $M(\piz\piz\piz)$ are determined from
MC simulations.

The fit to $M(\pip\pim\piz)$ and $M(\piz\piz\piz)$ distributions
yields the values $M(f_1) = 1287.4\pm3.0$~MeV/$c^2$ and $\Gamma(f_1) =
18.3\pm6.3$~MeV/$c^2$, with the number of events $N=78.2\pm19.3$ for
the $\Kp\Km\pip\pim\piz$ final state and $N=8.7\pm6.8$ ($<18.2$ at the
90\% Confidence Level (C. L.)) for the $\Kp\Km\piz\piz\piz$ final
state. The mass and width are consistent with those of the
axial-vector meson $f_1$~\cite{PDG14}~\footnote{Here we assume that the
  contribution of the pseudoscalar $\eta(1295)$ is small as no
  significant $\eta(1295)$ signals were found in the $\pip\pim\eta$
  mass spectrum from a study of $\jpsi \to
  \phi\pip\pim\eta$~\cite{songxy}.}. The statistical significance of
the $f_1$ signal is $5.2\sigma$ for the $\Kp\Km\pip\pim\piz$ final
state and $1.8\sigma$ for the $\Kp\Km\piz\piz\piz$ final state. From
the fit results, summarized in Table~\ref{tab:Nobs}, it is clear that
the production of a single $f_1$ resonance cannot account for all of
the $f_0\piz$ events above the background.

\section{Signal extraction of $\jpsi \to \phi \eta'$}
For the decay $\jpsi \to \phi \eta' \to \Kp\Km\pip\pim\piz$, the
decays $\jpsi \to \phi\eta' \to \Kp\Km\gamma \rho [(\gamma)\pip\pim]$
and $\jpsi \to \phi\eta' \to \Kp\Km\gamma\omega[\pip\pim\piz]$ produce
peaking background. To reduce the former peaking background which is
dominant, events with $0.920<M(\gamma\pip\pim)<0.970$~GeV/$c^2$ are
rejected.

As the amount of background for the decay
$\jpsi \to \phi \eta' \to\Kp\Km\piz\piz\piz$ is relatively small, the
$\phi$ signal and sideband regions are expanded to be
$1.010<M(\Kp\Km)<1.030$~GeV/$c^2$ and $1.040 <M(\Kp\Km)<1.060$~GeV/$c^2$,
respectively. A peaking background for this decay comes from the decay
$\jpsi \to \phi \eta'\to \Kp\Km\piz\piz\eta[\gamma\gamma]$. To reduce
this background, events with any photon pair mass in the range
$0.510<M(\gamma\gamma)<0.580$~GeV/$c^2$ are rejected.

\begin{figure}
  \subfigure{\includegraphics[width=0.4\textwidth]{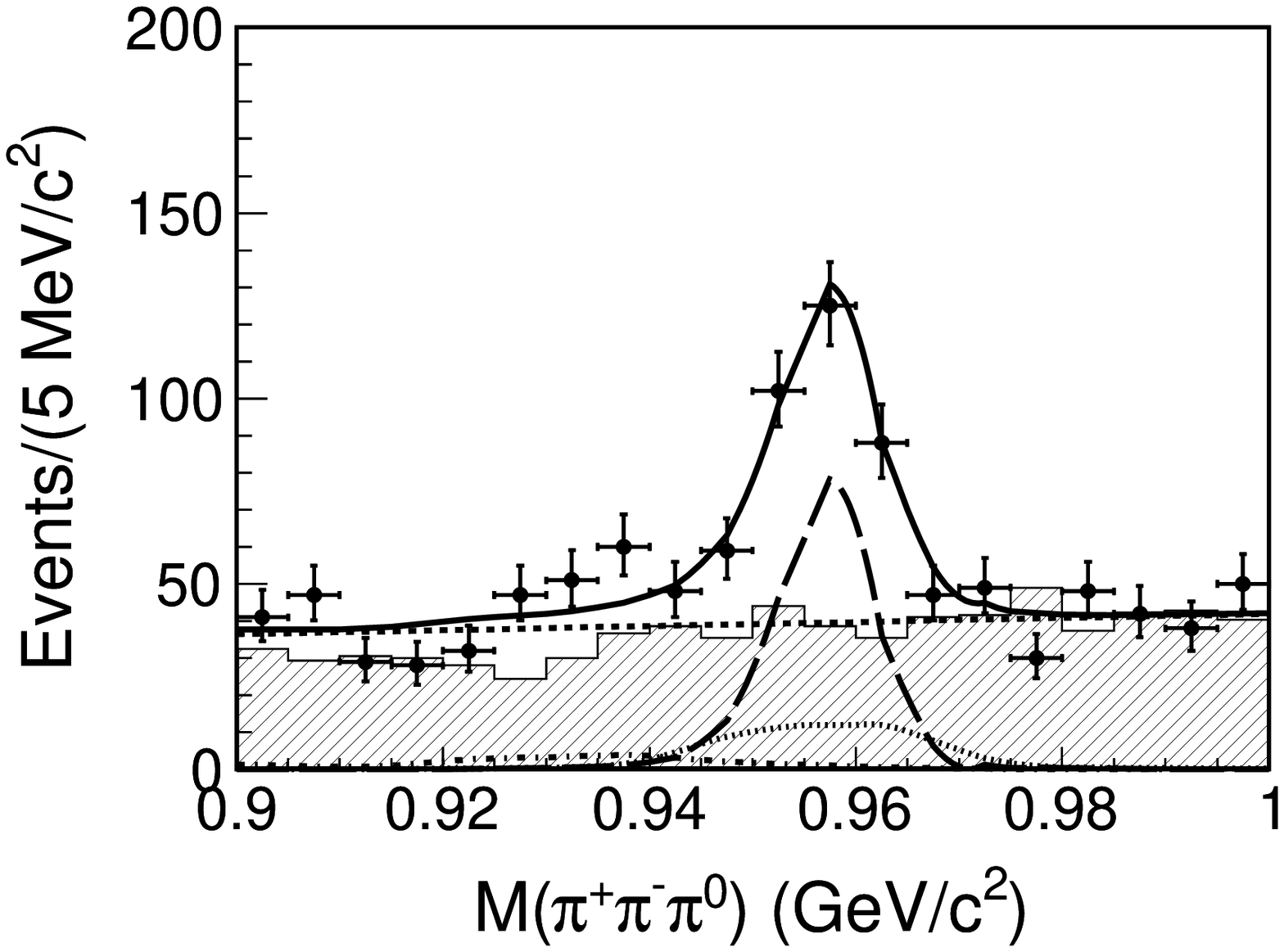}}
  \put(-155, 120){\textbf{(a)}}\\
  \subfigure{\includegraphics[width=0.4\textwidth]{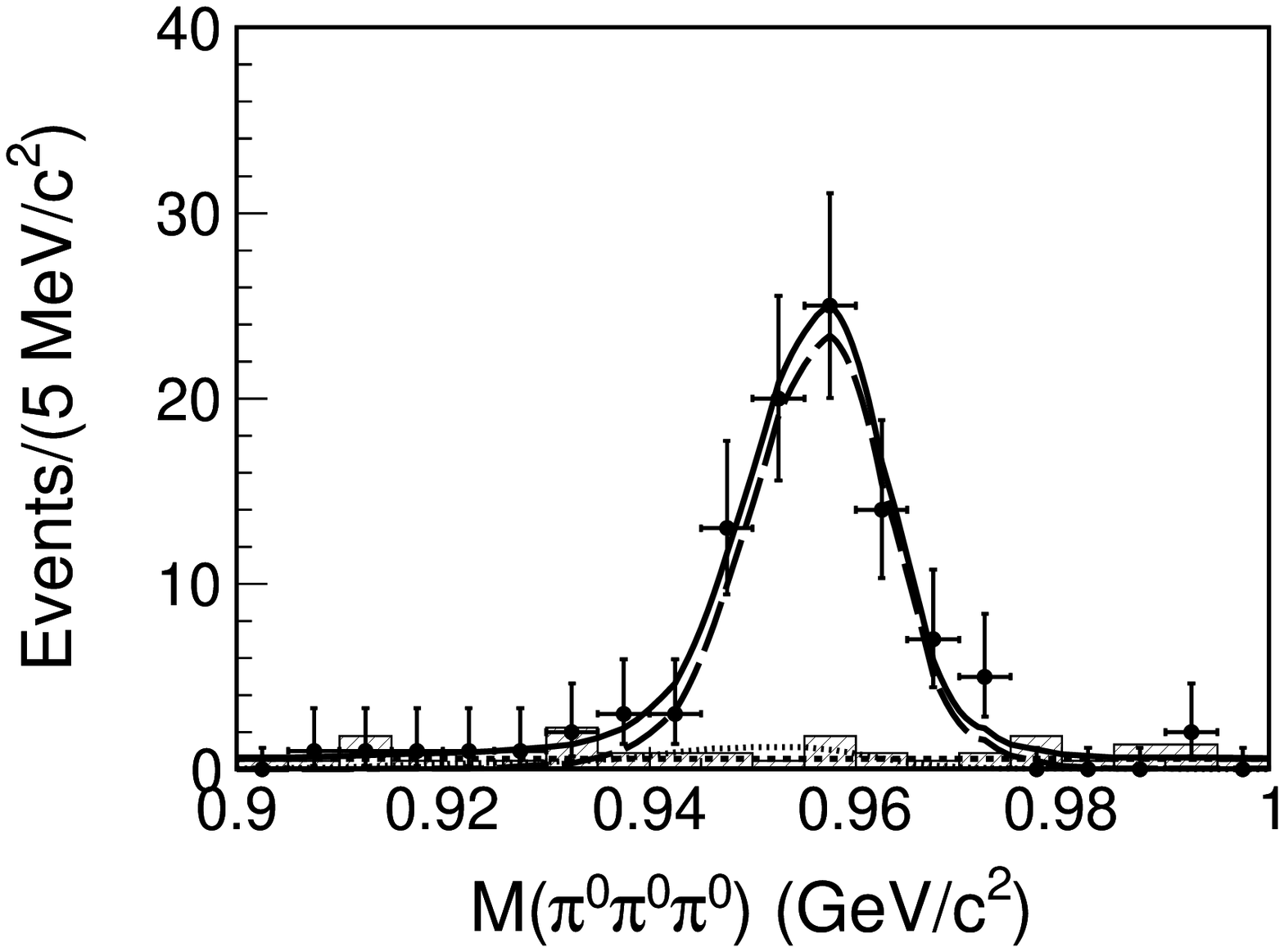}}
  \put(-155, 120){\textbf{(b)}}
  \caption{The spectra (a) $M(\pip\pim\piz)$ and (b) $M(\piz\piz\piz)$
    with $\Kp\Km$ in the $\phi$ signal region (the black dots) and
    sideband regions (the hatched histogram). The solid curve is the
    result of the fit, the long-dashed curve is the $\eta'$ signal,
    and the short-dashed line is the polynomial background. In (a),
    the dotted and dot-dashed curves represent the peaking background
    $\eta'\to \gamma\rho\to\gamma(\gamma)\pip\pim$ and
    $\eta' \to \gamma \omega \to \gamma \pip\pim\piz$,
    respectively. In (b), the dotted curve represents the peaking
    background $\eta' \to \piz\piz\eta$ with
    $\eta\to\gamma\gamma$.}\label{fig:etap}
\end{figure}

Figures~\ref{fig:etap} (a) and (b) show the final $\pip\pim\piz$ and
$\piz\piz\piz$ mass spectra for the $\phi$ signal (the black dots) and
sideband (the hatched histogram) regions. By analyzing data in the
$\phi$ sideband regions and the inclusive MC sample, we find that the
contribution from the decay $\jpsi \to \Kp\Km\eta'$ is negligible.

An unbinned likelihood fit is performed to obtain the signal
yields. The $\eta'$ signal shape is determined by sampling a histogram
from an MC simulation convoluted with a Gaussian function to compensate
for the resolution difference between the data and the MC sample. The
shape of the peaking background is determined from exclusive MC
samples, where the relative size of the background shape is determined
using the relevant branching fractions in the PDG~\cite{PDG14}. The
non-peaking background is described by a first-order (zeroth-order)
polynomial for the $\eta' \to\pip\pim\piz$ ($\piz\piz\piz$) decay. The
number of events are determined to be $N=183.3\pm 21.0$ for the
$\Kp\Km\pip\pim\piz$ final state and $77.6\pm9.6$ for the
$\Kp\Km\piz\piz\piz$ final state.

\begin{table}
\center{}
\caption{\label{tab:Nobs} Summary of the observed number of events
  ($N^\text{obs}$, the errors are statistical only.).}
\begin{tabular}{l l}
\hline\hline
Decay mode & $N^\text{obs}$\\
\hline
$\jpsi \to \phi\piz f_0$, $f_0 \to \pip\pim$ & $354.7\pm63.3$ \\
$\jpsi \to \phi\piz f_0$, $f_0 \to \piz\piz$ & $69.8\pm21.1$ \\
$\jpsi \to \phi f_1$, $f_1 \to \piz f_0$,$f_0 \to \pip\pim$ & $78.2\pm19.3$ \\
\multirow{2}{*}{$\jpsi \to \phi f_1$, $f_1 \to \piz f_0$, $f_0 \to \piz\piz$}  & $8.7\pm6.8$\\
& $< 18.2$ (90\% C.L.) \\
\hline
$\jpsi \to \phi\eta'$, $\eta' \to \pip\pim\piz$ & $183.3 \pm 21.0$\\
$\jpsi \to \phi\eta'$, $\eta' \to \piz\piz\piz$ & $77.6\pm9.6$\\
\hline\hline
\end{tabular}
\end{table}

\section{Branching fractions measurement}
Table~\ref{tab:Nobs} summarizes the signal yields extracted from the
fits for each decay. Equations~(\ref{eq:BRf0}) and (\ref{eq:BRetap})
give the formulae used to calculate the branching fractions, where $n$
is the number of $\piz$s in the final state $X$. $N^\text{obs}$ and
$\epsilon$ are the signal yield from the fits and efficiency from the
MC simulation for each decay, respectively. $B_{YZ}^X$ is the
branching fraction of the decay $X\to YZ$. $N_{\jpsi}$ is the number
of $\jpsi$ events. The upper limit of
$\BR(\jpsi \to \phi f_1, f_1 \to \piz f_0, f_0\to\piz\piz)$ is
determined according to Eq.~(\ref{eq:BRf1}), where $N_\text{upp}^\text{obs}$ is
the signal yield at the $90\%$ C. L. and $\sigma^{sys}$ is the total
systematic uncertainty, which is described in the next
section. Equation (\ref{eq:r3pi}) is used to calculate the ratio between
the branching fraction for $\eta' \to \piz\piz\piz$ and that for
$\eta' \to \pip\pim\piz$.

\begin{equation}
  \BR(\jpsi \to \phi X) = \frac{N^\text{obs}}{N_{\jpsi} \epsilon B_{\Kp\Km}^{\phi} (B_{\gamma\gamma}^{\piz})^n} \label{eq:BRf0}
\end{equation}
\begin{equation}
  \BR(\eta' \to X) = \frac{N^\text{obs}}{N_{\jpsi} \epsilon B_{\phi\eta'}^{\jpsi} B_{\Kp\Km}^{\phi} (B_{\gamma\gamma}^{\piz})^n} \label{eq:BRetap}
\end{equation}
\begin{equation}
  \BR(\jpsi \to \phi X) < \frac{N_\text{upp}^\text{obs}}{N_{\jpsi} \epsilon B_{\Kp\Km}^{\phi} (B_{\gamma\gamma}^{\piz})^n (1-\sigma^\text{sys})}\label{eq:BRf1}
\end{equation}

\begin{eqnarray}{\label{eq:r3pi}}
r_{3\pi} &\equiv& \BR(\eta' \to \piz\piz\piz)/\BR(\eta' \to \pip\pim\piz) \nonumber\\ &=&\frac{N^\text{obs}(\piz\piz\piz)}{N^\text{obs}(\pip\pim\piz)}\frac{\epsilon(\pip\pim\piz)}{\epsilon(\piz\piz\piz)}\frac{1}{(B_{\gamma\gamma}^{\piz})^2}
\end{eqnarray}

\section{Estimation of The Systematic Uncertainties}
\begin{itemize}
\item[(1)]\emph{MDC tracking}: The tracking efficiency of kaon tracks is studied using a high purity
sample of $\jpsi \to K_S K\pi$ events. The tracking efficiency of the
low-momentum pion tracks is studied using a sample of
$\jpsi \to \pip\pim p\bar{p}$ while that of the high-momentum pion
tracks is studied using a high statistics sample of
$\jpsi \to \rho \pi$. The MC samples and data agree within $1\%$ for
each kaon or pion track.

\item[(2)] \emph{Photon detection}: The photon detection efficiency is studied
using a sample of $\jpsi \to \rho \pi$ events. The systematic
uncertainty for each photon is $1\%$~\cite{Tracking}.

\item[(3)]\emph{PID efficiency}: To study the PID efficiency for kaon tracks, we select a clean sample
of $\jpsi \to \phi\eta \to \Kp\Km\gamma\gamma$. The PID efficiency is
the ratio of the number of events with and without the PID requirement
for both kaon tracks. MC simulation is found to agree with data within
0.5\%.

\item[(4)]\emph{Kinematic fit}: The performance of the kinematic fit is studied using a sample
$\jpsi \to \phi\eta \to \Kp\Km \pip\pim\piz/\Kp\Km\piz\piz\piz$, which
has the same final states as the signal channel
$\jpsi \to\phi\piz f_0$ with $\phi\to \Kp\Km$ and
$f_0 \to \pip\pim/\piz\piz$. The control sample is selected without
using the kinematic constraints. We then apply the same kinematic
constraints and the same requirement on the $\chi^2$ from the
kinematic fit. The efficiency is the ratio of the yields with and
without the kinematic fit. It contributes a systematic uncertainty of
1.0\% for $f_0 \to \pip\pim$ and 2.0\% for $f_0 \to \piz\piz$.

\item[(5)]\emph{Veto neutral $K^*$}: In selecting the candidate events
$\jpsi \to \phi \piz f_0 \to \Kp\Km\pip\pim\piz$, the events with
$|M(K^{\pm}\pi^{\mp})-M(K^{*0})|<0.050$ GeV/$c^2$ are vetoed to
suppress the background containing $K^{*0}$ or $\bar{K}^{*0}$
intermediate states. The requirement is investigated using a clean
sample $\jpsi \to \phi\eta \to \Kp\Km\pip\pim\piz$. The efficiency is
given by the yield ratio with and without the requirement
$|M(K^{\pm}\pi^{\mp})-M(K^{*0})|<0.050$ GeV/$c^2$. The efficiency
difference between data and MC is 0.1\%.

\item[(6)]\emph{$\phi$ signal region}: The uncertainty due to the
restriction on the $\phi$ signal region is studied with a high purity
sample of $\jpsi \to \phi \eta' \to \Kp\Km \pip\pim \eta$ events as
this sample is free of the background $\jpsi \to \Kp\Km\eta'$ without
the intermediate state $\phi$.

\item[(7)] \emph{Veto peaking background}: The uncertainties due to the
restrictions used to remove peaking background in the mode
$\eta' \to 3\pi$ are studied with a control sample of
$\jpsi \to \omega \eta \to 2(\pip\pim\piz)$ events. For each sample,
the efficiency is estimated by comparing the yields with and without
the corresponding requirement. The difference in efficiency between
the data and MC samples is taken as the systematic uncertainty.

\item[(8)]\emph{Background shape}: To study the effect of the
background shape, the fits are repeated with a different fit range or
polynomial order. The largest difference in signal yield is taken as
the systematic uncertainty.

\item[(9)]\emph{Mass resolution}: The mass resolutions, $\sigma_{MC}$, from an MC simulation of the modes
$f_0 \to \pip\pim/\piz\piz$ and $f_1 \to \piz f_0$ have an associated
systematic uncertainty. The difference in mass resolution, $\sigma_G$,
between the data and the MC simulation is determined using a sample of
$\jpsi \to \phi\eta$ events where
$\eta \to \pip\pim\piz/\piz\piz\piz$. The fit is repeated using
different mass resolutions, which are defined as
$\sqrt{\sigma_{MC}^2 + \sigma_G^2}$ assuming $\sigma_G$ is the same
for the two-pion and three-pion mass spectra. The difference in yield
is taken as a systematic uncertainty.

\item[(10)]\emph{MC simulation}: For the decay $\jpsi \to \phi \piz f_0$, the dominant systematic
uncertainty is from the efficiency $\epsilon_0$ determined by a phase
space MC simulation. The $\piz f_0$ invariant mass spectrum is divided
into 5 bins, each with a bin width of 0.2~GeV/$c^2$. The $f_0$ signal
yields, $N_i$, are determined by fits to the $\pi\pi$ spectra for each
bin $i$ using the mass and width of the $f_0$ obtained above. The
corrected efficiency is
$\epsilon_M \equiv \frac{\sum_iN_i}{\sum_i N_i/\epsilon_i}$, where
$\epsilon_i$ is the efficiency in the $i$-th bin. The same procedure
is applied to the angular distribution of the $\piz f_0$ system in the
c.m.~frame of the $\jpsi$ to obtain another corrected efficiency
$\epsilon_\theta$. The difference
$\sqrt{(\epsilon_M-\epsilon_0)^2 + (\epsilon_\theta-\epsilon_0)^2}$ is
taken as the systematic uncertainty due to the imperfection of the MC
simulation.

\item[(11)]\emph{$f_0$ signal region}: For the decay $\jpsi \to \phi f_1$ with $f_1 \to \piz f_0$, the $f_0$
signal region is $0.960<M(\pi\pi)<1.020$~GeV/$c^2$. The branching
fraction measurements are repeated after varying this region to
$0.970<M(\pi\pi)<1.010$~GeV/$c^2$ and
$0.950<M(\pi\pi)<1.030$~GeV/$c^2$. The differences from the nominal
results are taken as the systematic uncertainties due to the signal
region of the $f_0$. For the decay $f_1 \to \piz\piz\piz$, the number
of the peaking background $f_1\to\piz\piz\eta[\gamma\gamma]$ is
determined to be $3.1\pm0.6$. Varying the number of the peaking
background within $\pm0.6$ in the fit, the largest difference of the
signal yield gives a systematic uncertainty. The systematic
uncertainty values related to the $f_1$ are shown in brackets in
Table~\ref{tab:error}.

\item[(12)]\emph{About $\BR(\jpsi \to \phi f_1, f_1 \to \piz f_0, f_0 \to \piz\piz)$}: For the decay $\jpsi \to \phi f_1, f_1 \to \piz f_0$ with
$f_0 \to \piz \piz$, the signal yield at the $90\%$ C. L.,
$N_\text{upp}^\text{obs}$ in Eq.~(\ref{eq:BRf1}), is the largest one among the
cases with varying the fit ranges, the order of the polynomial
describing the background, the number of the peaking background, and
the signal region of the $f_0$ resonance. The total systematic
uncertainty, $\sigma^{sys}$ in Eq.~(\ref{eq:BRf1}), is the quadratic
sum of the rest systematic uncertainties in the third column of
Table~\ref{tab:error} (the values in the brackets). We obtain
$N_\text{upp}^\text{obs} = 29.0$ and $\sigma^{sys} = 6.9\%$ with the efficiency
$(7.21 \pm 0.08)\%$, determined from an MC simulation.
$\BR(\jpsi \to \phi f_1, f_1 \to \piz f_0, f_0 \to \piz\piz)$ is
calculated to be less than $6.98 \times 10^{-7}$ at the 90\%
C. L. according to Eq.~(\ref{eq:BRf1}).

\item[(13)]\emph{Uncertainty of $\BR(\jpsi \to \phi \eta')$}: For the decay $\eta' \to 3\pi$, the dominant systematic uncertainty
arises from the uncertainty of
$\BR(\jpsi \to \phi \eta')= (4.0\pm 0.7)\times
10^{-4}$~\cite{PDG14}.
A variation in $\BR(\jpsi \to \phi \eta')$ will change the size of
peaking background and thus the signal yield. In
Eq.~(\ref{eq:BRetap}), it is reasonable to consider a change in the
quantity $N^\text{obs}/B_{\phi\eta'}^{\jpsi}$ with any variation in
$\BR(\jpsi \to \phi \eta')$. The fit to the data is repeated after
varying the number of peaking background to correspond with $1\sigma$
variations in $\BR(\jpsi \to \phi \eta')$~\cite{PDG14}. The largest
difference of $N^\text{obs}/B_{\phi\eta'}^{\jpsi}$ from the nominal result
is taken as the systematic uncertainty.

\item[(14)]\emph{Systematic uncertainties for $r_{3\pi}$}: In the measurement of the ratio $r_{3\pi}$ of
$\BR(\eta' \to \piz\piz\piz)$ over $\BR(\eta' \to \pip\pim\piz)$, the
systematic uncertainties due to the reconstruction and identification
of kaon tracks and photon detection cancel as the efficiency ratio
$\epsilon(\piz\piz\piz)/\epsilon(\pip\pim\piz)$ appears in
Eq.~(\ref{eq:r3pi}). The effect of the uncertainty in the number of
peaking background due to the uncertainty of
$\BR(\jpsi \to \phi \eta')$ is also considered.

\end{itemize}
All systematic uncertainties including those on the number of $\jpsi$
events~\cite{Njpsi12} and other relevant branching fractions from the
PDG~\cite{PDG14} are summarized in Table~\ref{tab:error}, where the
total systematic uncertainty is the quadratic sum of the individual
contributions, assuming they are independent. Efficiency and branching
fraction measurements are summarized in Table~\ref{tab:BR}.

\begin{table*}
\center{}
\caption{\label{tab:error} Summary of systematic uncertainties ($\%$). For $f_0 \to \pi\pi$, the values in the brackets are for the decay $f_1 \to \piz f_0$. For $\eta' \to 3\pi$, the systematic uncertainty from the uncertainty of $\BR(\jpsi \to \phi \eta')$ is not included in the total quadratic sum. The last column lists the systematic uncertainties for the ratio between $\BR(\eta' \to \piz\piz\piz)$ and $\BR(\eta' \to \pip\pim\piz)$, denoted by $r_{3\pi}$.}
\begin{tabular}{l c c c c c}
\hline\hline
Sources &  $f_0 \to \pip\pim $ & $f_0 \to \piz\piz$ & $\eta' \to \pip\pim\piz$ & $\eta' \to 3\piz$ & $r_{3\pi}$\\
\hline
MDC tracking & 4.0 & 2.0 &  4.0 & 2.0 & 2.0\\
Photon detection & 2.0 & 6.0 & 2.0 & 6.0 & 4.0\\
PID efficiency& 0.5 & 0.5 & 0.5 & 0.5 & -\\
Kinematic fit & 1.0 & 2.0 &1.0 & 2.0 & 1.5\\
Veto neutral $K^{*}$ & 0.1 & -& -&- & -\\
$\phi$ signal region & 1.1 & 1.1 & 1.1 & 0.5 & 0.5\\
Veto peaking bkg. & - &- & 0.3 & 0.9 & 0.9 \\
Bkg. shape & 5.4 (15.5) & 4.4 (15.6) & 1.3 & 0.3 & 1.4\\
Mass resolution & 0.3 (0.4) & 1.0 (0.1) & - & - & -\\
MC simulation & 11.4 (-) & 11.4 (-) & - & - & -\\
$f_0$ signal region & -(2.4) & -(68.2) & - & - & - \\
$\BR(\jpsi \to \phi\eta')$ & - & - & 25.6 & 22.8 & - \\
Peaking bkg. &- &-(6.9) & - & - & 2.2\\
Number of $\jpsi$ &0.8 &0.8 & 0.8 & 0.8 &-\\
Other B.F. & 1.0 & 1.0 & 1.0 & 1.0 & 0.1\\
\hline
Total & 13.6 (16.5)  & 14.5 (70.6)& 5.1 & 6.9 & 5.5 \\
\hline\hline
\end{tabular}
\end{table*}

\begin{table*}
\center{}
\caption{\label{tab:BR} Summary of the efficiencies and the branching
  fractions. For the branching fractions, the first error indicates
  the statistical error and the second the systematic error. For
  $\BR(\eta'\to 3\pi)$, the third error is due to the uncertainty of
  $\BR(\jpsi \to \phi\eta')$~\cite{PDG14}. The last line gives the
  measured value of $r_{3\pi}$, defined as $\BR(\eta' \to
  \piz\piz\piz)/\BR(\eta'\to \pip\pim\piz)$.}
\begin{tabular}{l l l}
\hline\hline
Decay mode  & Efficiency ($\%$) & Branching fractions\\
\hline
$\jpsi \to \phi\piz f_0, f_0 \to \pip\pim$& $12.44 \pm 0.10$ & $(4.50\pm0.80\pm0.61)\times10^{-6}$\\
$\jpsi \to \phi\piz f_0, f_0 \to \piz\piz$  & $6.76 \pm 0.08$ & $(1.67\pm0.50\pm0.24)\times10^{-6}$\\
$\jpsi \to \phi f_1, f_1 \to \piz f_0 \to \piz\pip\pim$  & $13.19 \pm 0.11$ & $(9.36 \pm 2.31 \pm 1.54)\times10^{-7}$\\
$\jpsi \to \phi f_1, f_1 \to \piz f_0 \to \piz\piz\piz$  & $6.76 \pm 0.08$ & $(2.08 \pm 1.63 \pm 1.47)\times10^{-7}$\\
& & $<6.98\times10^{-7}$ ($90\%$ C. L.)\\
\hline
$\eta' \to \pip\pim\piz$ & $16.92 \pm 0.12$ & $(4.28 \pm 0.49 \pm 0.22 \pm 1.09)\times10^{-3}$ \\
$\eta' \to \piz\piz\piz$ & $6.55 \pm 0.08$ & $(4.79 \pm0.59\pm0.33\pm 1.09)\times10^{-3}$\\
\hline
$r_{3\pi}$ & & $1.12\pm0.19\pm0.06$ \\
\hline\hline
\end{tabular}
\end{table*}

\section{Summary}
In summary, we have studied the decay
$\jpsi \to \phi 3\pi \to \Kp\Km 3\pi$. The isospin violating decay
$\jpsi \to \phi \piz f_0$ is observed for the first time. In the
$\piz f_0$ mass spectrum, there is an evidence of the axial-vector
meson $f_1$, but not all $\piz f_0$ pairs come from the decay of an
$f_1$. Using $\BR(\jpsi \to \phi f_1) = (2.6\pm0.5) \times 10^{-4}$
and $\BR(f_1 \to \pi a_0 \to \pi\pi\eta)=(36\pm7)\%$ from the
PDG~\cite{PDG14}, the ratio
$\BR(f_1 \to \piz f_0 \to \piz\pip\pim)/\BR(f_1 \to \piz a_0^0\to
\piz\piz\eta)$
is determined to be $(3.6\pm1.4)\%$ assuming isospin symmetry in
the decay $f_1 \to a_0\pi$. This value is only about $1/5$ of
$\BR(\eta(1405) \to \piz f_0 \to \piz\pip\pim)/\BR(\eta(1405) \to \piz
a_0^0\to \piz\piz\eta) = (17.9 \pm 4.2)\%$~\cite{BESgpi0f0}.
On the other hand, the measured mass and width of the $f_0$ obtained
from the invariant dipion mass spectrum are consistent with those in
the decay
$\jpsi \to \gamma \eta(1405) \to \gamma\piz f_0$~\cite{BESgpi0f0}. The
measured $f_0$ width is much narrower than the world average value of
$40-100$~MeV~\cite{PDG14}.
It seems that there is a contradiction in the isospin-violating decays $f_1/\eta(1405) \to \piz f_0$.
However, a recent theoretical work~\cite{f1f0}, based on the triangle singularity mechanism as proposed in Ref.~\cite{JJWu,Aceti}, analyzes the decay $f_1 \to \piz f_0 \to \piz \pip\pim$ and predicts that the width of the peaking structure in the $f_0$ region is about 10 MeV. It also derives $\BR(f_1 \to \piz f_0 \to \piz\pip\pim)/\BR(f_1 \to \piz a_0^0\to
\piz\piz\eta) \simeq 1\%$, which is close to our measurement. This analysis supports the argument that the nature of the resonances $a_0^0$ and $f_0$ as dynamically generated makes the amount of isospin breaking strongly dependent on the physical process~\cite{f1f0}.
In addition, we have measured the branching fractions
$\BR(\eta' \to \pip\pim\piz) = (4.28 \pm 0.49 (\text{stat.}) \pm 0.22(\text{syst.})
\pm 1.09) \times 10^{-3}$
and
$\BR(\eta' \to \piz\piz\piz) = (4.79 \pm0.59(\text{stat.})\pm0.33(\text{syst.})\pm
1.09) \times 10^{-3}$,
where the last uncertainty is due to $\BR(\jpsi \to \phi \eta')$. The
ratio between them $r_{3\pi} = 1.12 \pm 0.19(\text{stat.}) \pm 0.06(\text{syst.})$
is also measured for the first time. These results are consistent with
those measured in the decay $\jpsi \to \gamma\eta'$~\cite{BESgpi0f0}.

\section{Acknowledgement}
The BESIII collaboration thanks the staff of BEPCII and the IHEP computing center for their strong support. This work is supported in part by National Key Basic Research Program of China under Contract No. 2015CB856700; National Natural Science Foundation of China (NSFC) under Contracts Nos. 11125525, 11235011, 11322544, 11335008, 11425524; the Chinese Academy of Sciences (CAS) Large-Scale Scientific Facility Program; the CAS Center for Excellence in Particle Physics (CCEPP); the Collaborative Innovation Center for Particles and Interactions (CICPI); Joint Large-Scale Scientific Facility Funds of the NSFC and CAS under Contracts Nos. 11179007, U1232201, U1332201; CAS under Contracts Nos. KJCX2-YW-N29, KJCX2-YW-N45; 100 Talents Program of CAS; INPAC and Shanghai Key Laboratory for Particle Physics and Cosmology; German Research Foundation DFG under Contract No. Collaborative Research Center CRC-1044; Istituto Nazionale di Fisica Nucleare, Italy; Ministry of Development of Turkey under Contract No. DPT2006K-120470; Russian Foundation for Basic Research under Contract No. 14-07-91152; U. S. Department of Energy under Contracts Nos. DE-FG02-04ER41291, DE-FG02-05ER41374, DE-FG02-94ER40823, DESC0010118; U.S. National Science Foundation; University of Groningen (RuG) and the Helmholtzzentrum fuer Schwerionenforschung GmbH (GSI), Darmstadt; WCU Program of National Research Foundation of Korea under Contract No. R32-2008-000-10155-0

\end{document}